# Chiral Corrections to Lattice Calculations of Charge Radii

DEREK B. LEINWEBER AND THOMAS D. COHEN
*Department of Physics and Center for Theoretical Physics*
*University of Maryland, College Park, MD 20742*

November, 1992

Logarithmic divergences in pion and proton charge radii associated with chiral loops are investigated to assess systematic uncertainties in current lattice determinations of charge radii. The chiral corrections offer a possible solution to the long standing problem of why present lattice calculations yield proton and pion radii which are similar in size.

# I. INTRODUCTION

While it is generally accepted that quantum chromodynamics (QCD) is the correct theory of strong interactions and, as such, is sufficient to explain the structure of hadrons, it is extremely difficult to solve the theory in the low momentum transfer region. The most promising technique for eventually deriving low energy properties of hadrons directly from QCD is via numerical Monte Carlo simulations of the functional integral in lattice regularized QCD. The examination of hadron structure through lattice techniques has provided many new insights to our understanding of quark dynamics [1–3].

Unfortunately, given the current computational power and algorithms available, it is not possible to directly calculate hadron properties in a completely realistic manner. One important limitation is the present inability to directly calculate hadron properties with light $u$- and $d$-current-quark masses.

There is another approach to hadronic structure which in one important respect is complementary to lattice QCD. This approach, based on minimal model assumptions, is the expansion in $m_\pi^2$ of chiral perturbation theory ($\chi$PT) [4–7]. This expansion is motivated by a separation between the scale of the pion mass and other mass scales in the QCD spectrum. The approach becomes increasingly valid as the current-quark masses become light. In contrast, critical behavior in the quark mass makes the lattice approach increasingly intractable as quark masses become light.

In this paper, we will investigate the relationship between lattice QCD and $\chi$PT. Information from $\chi$PT might help one extrapolate lattice results obtained with rather heavy quark masses down to the physical point. Similarly, information obtained with lattice calculations may help one determine the range of validity for $\chi$PT estimates of various quantities.

The lightest quark mass parameters have been investigated in spectroscopy calculations where typical $\pi$ to $\rho$ mass ratios are found to be of order $1/2$ [8–10]. In an attempt to make some connection with the physical world where $m_\pi/m_\rho \simeq 0.18$, hadron properties are often extrapolated in quark mass to the chiral limit where the pion becomes massless. In general, given the values of quark mass presently used, the values of various quantities when extrapolated to the physical point (using simple linear extrapolations) are essentially indistinguishable from extrapolations to the chiral limit.

The principal idea we will explore in this paper is that $\chi$PT suggests corrections to the standard linear extrapolation of lattice results. Consider the role of "pion loops" in a hadronic description of various quantities of interest. The presence of the pion mass in the propagators can lead to infrared divergences in the chiral limit, $m_\pi \to 0$, and hence to *nonanalytic* behavior of quantities as a function of the pion mass. Of course, the meaning of a pion loop in QCD is somewhat obscure. However, the physics underlying the nonanalytic behavior is simply the existence of nearby singularities in a dispersion relation treatment of the quantities of interest.

In this paper we will focus on the electromagnetic charge radii of hadrons. These are particularly interesting since they have dramatic nonanalytic effects. $\chi$PT predicts that the electric charge radius of both the proton and the pion diverge as one approaches the chiral limit. This divergence is logarithmic in the pion mass. Moreover, these radii have recently been extracted in lattice calculations [1, 2]. By including these chiral logarithms in the extrapolations with respect to the quark mass, the systematic error associated with the present lattice calculations may be assessed.

## II. CHARGE RADII FROM LATTICE QCD

To investigate these effects, we will consider the lattice results of Ref. [1] and [2] for the pion and proton form factors. These lattice calculations are based on a numerical simulation of quenched QCD on a $24 \times 12 \times 12 \times 24$ lattice at $\beta = 5.9$ using Wilson fermions. Twenty-eight quenched gauge configurations are used in the analysis. Statistical uncertainties in the lattice results are determined using a third order, single elimination jackknife [11]. The form factors or radii are determined at three values of the Wilson hopping parameter, $\kappa$, and are extrapolated linearly in $1/\kappa$ to $1/\kappa_{cr}$ where the pion mass vanishes. Traditionally one uses a linear extrapolation in $1/\kappa$ (or $m_q$) of the squared pion mass to zero in determining the value of $\kappa_{cr}$. Thus the extrapolation of radii in $1/\kappa$ may be equivalently described as an extrapolation in $m_\pi^2$.

It is worth noting that in the extrapolation of the pion mass squared, the effects of terms higher order in $m_q$, expected in a $\chi$PT expansion (even in the quenched approximation [12, 13]), have been neglected in using a simple linear extrapolation. Errors in other extrapolated quantities such as charge radii will be induced by the neglect of these nonlinearities. However, these errors will be at most of order $m_q \ln(m_q)$ (*i.e. finite*) as one approaches the chiral limit. In contrast, the leading order chiral corrections to the charge radii diverge.

In the analyses of Ref. [1, 2] charge radii are determined by parameterizing the charge form factor to a dipole or monopole form, from which a charge radius may be inferred. Many different extrapolation methods were tested in determining the extrapolated values for the charge radii. In the following we define a conventional extrapolation scheme as fitting the lattice determination of the form factor to a dipole form, followed by an extrapolation of the radii to the chiral limit as a function of $1/\kappa$ or equivalently as a function of $m_\pi^2$.

For the pion, the momentum transfer varies by a factor of 2 over the range between the lattice calculations and the physical pion mass [1]. While an extrapolation of the pion form factor is unreliable, an extrapolation of the pion charge radius may be more reasonable. At $q^2 \simeq 0.15$ GeV$^2$ where the lattice form factors are determined, the dipole and monopole parameterizations of the form factor produce radii which agree within statistical uncertainties. This is due to the fact that the curvature associated with dipole and monopole forms is insignificant in this regime of $q^2$. A linear parameterization of the form factors also produces similar radii. Therefore the radius, proportional to the derivative of the form factor, is much less sensitive to variations in $q^2$.

The assumption in performing a linear extrapolation for the charge radii is that any significant curvature due to nonanalytic terms of $\chi$PT lies between the physical point and the chiral limit. The similarity of empirical isoscalar and isovector radii of the nucleon suggest that such an approximation is not unjustified. However in this paper we will take chiral corrections more seriously in an attempt to estimate the possible size of systematic errors in present calculations.

### III. CHIRAL CORRECTIONS

The logarithmic divergences in the pion and proton charge radii associated with chiral loops are well known [14]. The expressions, which become increasingly better as one approaches the chiral limit are

$$\langle r_\pi^2 \rangle = \frac{1}{(4\pi)^2} \frac{1}{f_\pi^2} \ln\left(\frac{q_0^2}{m_\pi^2}\right) + F_\pi(m_\pi) \qquad (3.1)$$

$$\langle r_p^2 \rangle = \frac{1}{2}\left(\frac{g_{\pi NN}^2}{4\pi} \frac{3}{2\pi M^2} + \frac{1}{(4\pi)^2} \frac{1-g_A^2}{f_\pi^2}\right) \ln\left(\frac{q_0^2}{m_\pi^2}\right)$$
$$+ F_p(m_\pi). \qquad (3.2)$$

Here $f_\pi$, $g_{\pi NN}$, $g_A$, and $M$ are the pion decay constant, the pion-nucleon coupling constant, the axial coupling constant and the nucleon mass respectively. The parameter $q_0$ is a constant with dimensions of mass. In the preceding expression $g_{\pi NN}$ refers to the pion-nucleon coupling constant at $q^2 = 0$. The Goldberger-Treiman relation is exact at $q^2 = 0$ and thus the quantity $g_{\pi NN}/M$ can be replaced by $g_A/f_\pi$. Moreover, $g_{\pi NN}$ and $M$ only enter the expression as a ratio and therefore to determine the singular parts only $g_A$ and $f_\pi$ need be specified.

In principle, these parameters should be evaluated in the chiral limit. On the other hand, these parameters correspond to observables which are free from singularities in the chiral limit. Thus, in the chiral limit, these quantities equal their physical values plus correction terms of order $m_\pi^2$. Such corrections have been estimated to be the order of a few percent [5, 6] and for our present qualitative purpose it is sufficient to take the physical values. We take $f_\pi = 93.2$ MeV and $g_A = 1.26$. Of course, along with the singular terms there are contributions which remain finite as the chiral limit is approached. This physics is accounted for in the functions $F_\pi$ and $F_p$.

The results of Ref. [1, 2] are obtained in the quenched approximation of QCD. When comparing $\chi$PT with lattice calculations, one may wish to address the question of whether the physics which dominates the chiral singularities (i.e. pion loops) is, in fact, present in the quenched approximation [12, 13, 15, 16]. However, this question only affects the interpretation of the results rather than the predictions.

If quenched calculations do have the correct chiral physics, then as will be shown in this paper, there may be large systematic errors in extrapolated quantities if one does not account for chiral logarithmic terms in the extrapolation. On the other hand, if the quenched approximation does not include the *correct* chiral behavior [17] then the inability to reproduce this behavior is itself an important systematic error of the quenched approximation. Of course, the assumption implicit in this approach is that the lattice results for charge radii at the relatively heavy quark masses considered in Ref. [1, 2] are similar to what one would extract in an analysis of full QCD. This assumption is generally supported by present hadronic mass spectrum analyses. However, it is not so apparent that the quenched approximation will remain a good approximation when the quark masses become light. In fact the discrepancies between full and quenched $\chi$PT to some extent indicate the breakdown of the quenched approximation for many quantities.

Alternatively one may wish to extrapolate the physical charge radii to the regime of quark masses or pion masses typically investigated in lattice calculations. To extend the $\chi$PT physics into this regime we consider the following assumption for the finite terms. The obvious and standard choice for $F(m_\pi)$ is

$$F(m_\pi) = C_0 + C_1 m_\pi^2. \qquad (3.3)$$

The contribution of the unknown constant $q_0$ may be absorbed into $C_0$ and therefore we define $q_0 = a^{-1}$ the inverse lattice spacing and consider the extrapolations

$$\langle r_\pi^2 \rangle = \frac{1}{(4\pi)^2} \frac{1}{f_\pi^2} \ln\left(\frac{1}{m_\pi^2 a^2}\right) + C_0^\pi + C_1^\pi m_\pi^2 \qquad (3.4)$$

$$\langle r_p^2 \rangle = \frac{1}{2} \frac{1}{(4\pi)^2} \frac{1}{f_\pi^2} \left(1 + 5g_A^2\right) \ln\left(\frac{1}{m_\pi^2 a^2}\right)$$
$$+ C_0^p + C_1^p m_\pi^2. \qquad (3.5)$$

The physical constants appearing in these equations are also functions of the pion mass. Variations in these parameters give rise to higher order terms in the expansion. To estimate the importance of these higher order terms we also consider extrapolations of the radii where $f_\pi$ is a function of $m_\pi$. The extraction of the $m_\pi$ dependence of $g_A$ on the lattice is complicated by a mass dependence in the renormalization constant. For present purposes it is sufficient to leave it fixed. For the $m_\pi$ dependence of $f_\pi$ we use

$$f_\pi(m_\pi^2) \sim f_\pi(0) + (0.081 \text{ GeV}^{-1}) m_\pi^2, \qquad (3.6)$$

where we have used the APE collaboration results [9] for $f_\pi$ to determine the coefficient of $m_\pi^2$. Results from a recent lattice simulation of full QCD suggest a similar dependence for $f_\pi$ [18].

We have not used $\chi$PT to estimate the $m_\pi^2$ dependence of $f_\pi$. Furthermore, using an $m_\pi^2$ dependence for $f_\pi$ is not consistent with leading order $\chi$PT. The fact is we do not know how to study higher order effects in a systematic way. We have simply included the $m_\pi^2$ dependence in $f_\pi$ in order to assess the scale of higher order terms which might appear in the expansion.

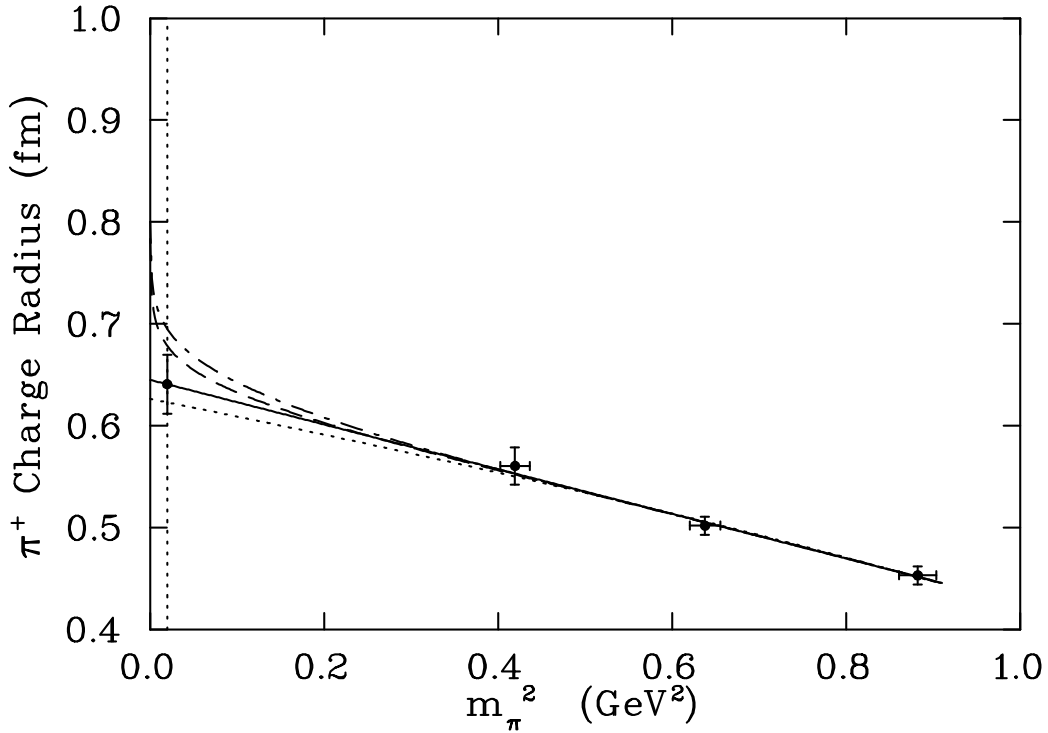

FIG. 1 Extrapolations of the lattice pion radii to the physical regime indicated by the vertical dotted line. In this and the following figure the axes have been scaled by a constant lattice spacing of 0.131 fm to provide a reference scale to otherwise dimensionless quantities. From top to bottom the extrapolation curves correspond to equation (3.4) with $f_\pi$ as in (3.6), (3.4) with $f_\pi$ fixed, the conventional linear extrapolation of $\langle r^2 \rangle^{1/2}$ and a linear extrapolation of $\langle r^2 \rangle$.

The radius of convergence in $m_\pi^2$ for (3.4) and (3.5) is unknown. Indeed, it is unknown if the series is convergent. However, the contributions of the logarithmic terms are largest when the pion mass is small and more likely to be the dominant physics. Conversely, at pion masses typical of present lattice calculations one might be in a regime beyond the validity of low order $\chi$PT.

The parameters $C_0$ and $C_1$ are optimized by fitting the lattice determinations of the radii at the 3 different values of pion mass by minimizing a chi-square measure weighted by the square of the lattice statistical uncertainties. Figures 1 and 2 display the results for the various extrapolation procedures. For both the pion and proton radii, it is clear that smaller quark masses will be required to reveal any logarithmic divergences in lattice QCD simulations. The plots also indicate an underestimation of the radii using conventional lattice extrapolation techniques. The pion radius is underestimated by approximately 5% and the proton radius by 25%. Some sensitivity to higher order terms in the chiral extrapolations is displayed in the difference between the curves with $f_\pi$ fixed and $f_\pi$ a function of $m_\pi^2$. In table I, charge radii obtained using linear extrapolations in $m_\pi^2$ or $m_\pi$ [19], and the standard chiral expansions of (3.4) and (3.5) with $f_\pi$ fixed are compared with experiment. It is amusing to note that the chiral correction to the proton charge radius is precisely the amount required to restore agreement between the lattice and experimental results.

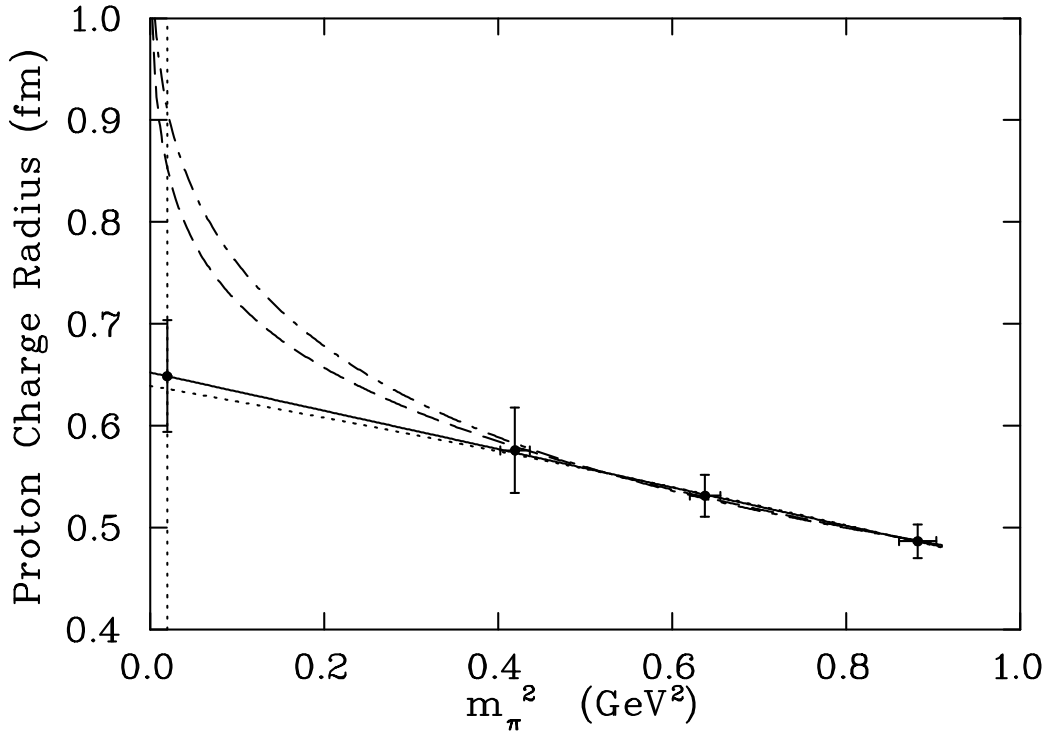

FIG. 2 Extrapolations of the lattice proton radii to the physical point. From top down, the extrapolation curves correspond to equation (3.5) with $f_\pi$ as in (3.6), (3.5) with $f_\pi$ fixed, the conventional linear extrapolation of $\langle r^2 \rangle^{1/2}$ and a linear extrapolation of $\langle r^2 \rangle$.

TABLE I. Comparison of extrapolated charge radii with experimental results. Radii are in units of fm.

| Hadron | Linear in $m_\pi^2$ | Linear in $m_\pi$ | $\chi$PT | Expt. |
|---|---|---|---|---|
| $\pi$ | 0.64(7) | 0.74(9) | 0.68(6) | 0.66 ±0.01 |
| $p$ | 0.65(5) | 0.73(7) | 0.88(3) | 0.862±0.012 |

## IV. DISCUSSION

We have shown the possibility of significant corrections to conventional linear extrapolations of charge radii. The inclusion of chiral logarithms enhances both the extrapolated pion and proton charge radii and offers a possible solution to the long standing problem of why present lattice calculations yield radii for these hadrons which are similar in size.

We do not claim that these "chirally corrected" predictions of charge radii are reliable. There are a number of reasons to question the validity of the additional terms. A critical point is that we have only included a single chiral correction term, although it is divergent in the chiral limit. $\chi$PT also allows the inclusion of terms proportional to $m_\pi^4$, $m_\pi^6$ etc. While it may be legitimate to conclude that such terms are negligible near the chiral limit (and hopefully up to the physical point), it is not *a priori* obvious that such terms are negligible in the region of quark mass where the lattice calculations are done.

There is a more fundamental question about $\chi$PT. The radius of convergence of $\chi$PT is unknown. If the series is asymptotic the number of terms which are reliable is also unknown. It is generally believed to be valid for physical quark masses but it has been suggested that the $\chi$PT approach may break down with quark masses as light as the scale of the strange quark mass [20]. There have been recent suggestions that the standard $\chi$PT approach to nucleon properties is not valid when the pion mass becomes of comparable scale to the N-$\Delta$ mass splitting [21].

We believe, however, that studies of these chiral corrections are significant. They give us some estimate of the scale of the errors associated with the current lattice QCD predictions of charge radii. Figures 1 and 2 give us an estimate of how important these chiral logarithms are near the physical point and hence how large a systematic error may be in any calculation which does not account for them. It is clear that to obtain accurate predictions of charge radii it is necessary to reduce the systematic errors in the quark mass extrapolations. This may not be possible until lattice calculations with much lighter quark masses become available. In the meson sector the predictions of quenched $\chi$PT [12, 13, 16] may be useful due to the large separation between the pion mass and all other relevant mass scales. In the baryon sector, the proximity of baryon excitations at the scale of the pion mass demands a more careful treatment perhaps including many baryon excitations in $\chi$PT.

We do not view the divergences anticipated by $\chi$PT as a problem insurmountable to lattice QCD methods. The logarithmic terms of the extrapolation function have an associated finite volume dependence [22] which may be examined to determine the coefficients of the logarithms. For the quenched approximation, techniques are under development for the calculation of the logarithmic coefficients anticipated by quenched $\chi$PT [12, 13, 16] which may be used to verify the physics included in the quenched approximation [16] and evaluate the validity of $\chi$PT itself. Until lattice calculations with realistic quark masses become feasible, one hopes to be able to use an extrapolation function motivated by generalized $\chi$PT considerations with all input parameters determined self consistently on the lattice in order to make contact with the physical world.

## ACKNOWLEDGMENTS

We are indebted to Terry Draper for providing the pion correlation functions used in this analysis. D.B.L. thanks Richard Woloshyn and Terry Draper for early discussions which stimulated his interest in these issues. This work is supported in part by the U.S. Department of Energy under grant DE-FG05-87ER-40322 and the National Science Foundation though grant PHY-9058487.